\documentclass[11pt]{article}
	
	\newcommand{\blind}{0}
    \newcommand{\beginsupplement}{%
        \setcounter{table}{0}
        \renewcommand{\thetable}{S\arabic{table}}%
        \setcounter{figure}{0}
        \renewcommand{\thefigure}{S\arabic{figure}}%
     }
    \makeatletter
    \renewcommand\section{\@startsection {section}{1}{\z@}%
                                       {-3.5ex \@plus -1ex \@minus -.2ex}%
                                       {2.3ex \@plus.2ex}%
                                       {\normalfont\fontfamily{phv}\fontsize{16}{19}\bfseries}}
    \renewcommand\subsection{\@startsection{subsection}{2}{\z@}%
                                         {-3.25ex\@plus -1ex \@minus -.2ex}%
                                         {1.5ex \@plus .2ex}%
                                         {\normalfont\fontfamily{phv}\fontsize{14}{17}\bfseries}}
    \renewcommand\subsubsection{\@startsection{subsubsection}{3}{\z@}%
                                        {-3.25ex\@plus -1ex \@minus -.2ex}%
                                         {1.5ex \@plus .2ex}%
                                         {\normalfont\normalsize\fontfamily{phv}\fontsize{14}{17}\selectfont}}
    \makeatother
	
	\usepackage{amsmath}
	\usepackage{graphicx}
	\usepackage{enumerate}
	\usepackage{xcolor}
	\usepackage{url}
  \usepackage[colorinlistoftodos]{todonotes} 
  \usepackage{longtable}
  \usepackage{lineno}
  \usepackage{floatrow}
  \usepackage{float}
  \usepackage{caption}
  \usepackage{booktabs}
  \usepackage{multirow}
  \usepackage{xcolor}
  \usepackage[style=apa]{biblatex}
  \usepackage{comment}
  \usepackage{xr} 
  \floatstyle{plaintop}
  \restylefloat{table}

\begin{document}

		\def\spacingset#1{\renewcommand{\baselinestretch}%
			{#1}\small\normalsize} \spacingset{1}
		
		\if0\blind
		{
			\title{\bf Can machine learning predict citizen-reported angler behavior?}
			
			\author{Julia S. Schmid\thanks{Corresponding author (email: jschmid@ualberta.ca)} $^1$, Sean Simmons$^2$, Mark A. Lewis$^{1,3,4,5}$, Mark S. Poesch$^5$, Pouria Ramazi$^6$ \\
			\small
			   $^1$Department of Mathematical and Statistical Sciences, University of Alberta, Edmonton, Alberta, Canada \\
			 \small
             $^2$Angler’s Atlas, Goldstream Publishing, Prince George, British Columbia, Canada \\
             \small
             $^3$Department of Mathematics and Statistics, University of Victoria, Victoria, British Columbia, Canada \\
             \small
             $^4$Department of Biology, University of Victoria, Victoria, British Columbia, Canada \\
             \small
             $^5$Department of Biological Sciences, University of Alberta, Edmonton, Alberta, Canada \\
             \small
             $^6$Department of Mathematics and Statistics, Brock University, St. Catharines, Ontario, Canada}
			\date{}
			\maketitle
		} \fi

\newpage
	\begin{abstract}
    Prediction of angler behaviors, such as catch rates and angler pressure, is essential to maintaining fish populations and ensuring angler satisfaction. Angler behavior can partly be tracked by online platforms and mobile phone applications that provide fishing activities reported by recreational anglers. Moreover, angler behavior is known to be driven by local site attributes. Here, the prediction of citizen-reported angler behavior was investigated by machine-learning methods using auxiliary data on the environment, socioeconomics, fisheries management objectives, and events at a freshwater body. The goal was to determine whether auxiliary data alone could predict the reported behavior. Different spatial and temporal extents and temporal resolutions were considered. Accuracy scores averaged 88\% for monthly predictions at single water bodies and 86\% for spatial predictions on a day in a specific region across Canada. At other resolutions and scales, the models only achieved low prediction accuracy of around 60\%. The study represents a first attempt at predicting angler behavior in time and space at a large scale and establishes a foundation for potential future expansions in various directions.
  \end{abstract}
			
	\noindent%

	\spacingset{1} 

\newpage
\section{Introduction} \label{s:intro}

Fishing has a high value in our society by providing multiple services such as recreation, food and social interaction (\cite{Fedler1994}).
In order to prevent overfishing and maintain a healthy fishery, the assessment of angler behavior and fishery dynamics is an essential element of precautionary fisheries management strategies (\cite{Post2002, Hunt2011, Bade2022}).

Machine-learning methods and statistical models are increasingly used for monitoring, predicting, and understanding recreational angler pressure and catch rates due to their potential to uncover intricate patterns (\cite{Kaemingk2020, Sasamal2019, Behivoke2021, Powers2016, Heermann2013}). For instance, monthly recreational marine fishing effort was predicted into the future at two fisheries using boosted regression trees based on fishing logbook records, and temporal and environmental features explained 58.7\% to 65.6\% of the deviance in the data (\cite{Soykan2014}). Decisions of angler on harvest-release were correctly predicted for 96\% of fish caught using extreme gradient boosted decision trees based on creel survey data including social and ecological variables (\cite{Kaemingk2019, Kaemingk2020}). 

However, studies on spatial and temporal predictions of angler pressure and catch rates were usually limited to a few water bodies or a specific region (\cite{Soykan2014, Behivoke2021, Powers2016}). Data from conventional angler surveys, like in-person interviews, that are used for model training are very costly and limited in space and time. In addition, the methodologies of these surveys can differ in terms of measuring angler pressure and the specific survey methods used in different locations and at different times, making it difficult to combine them on a large scale.

An alternative data source that can cover large and fine spatial extents and temporal resolutions are online platforms or mobile phone applications (apps). Recreational anglers can use apps to report their trips to track and evaluate their fishing success, and get information on specific water bodies (\cite{Jiorle2016, Gundelund2020, Johnston2022}). The apps can collect different information like the fishing location, the date and duration of the trip, and the catch rate. Although this citizen-reported data can be biased as it only represents the angler behavior of a certain group of anglers that use the platform, it can be consistent with conventional survey methods and lead to similar conclusions (\cite{Johnston2022, Martin2014, Gundelund2020, Gundelund2021CJFAS, Papenfuss2015}). For instance, estimates of catch rates based on citizen-reported app data were found to be similar to catch rates derived from surveys in Alberta and Florida (\cite{Johnston2022, Jiorle2016}). \textcite{Papenfuss2015} found consistencies between app and conventional data regarding travel distances and temporal patterns of angler visits at lakes in Alberta.  Moreover, it has been shown that such data can be useful to investigate the spread of fish diseases (\cite{Fischer2021, Ramazi2021EE}) or invasive species (\cite{Papenfuss2015, Weir2022}).

Here, an array of different machine-learning methods together with auxiliary data on the environment, socioeconomics, and fisheries management and events was used to make spatial and temporal predictions of app-reported angler behavior across freshwater bodies in Canada. More specifically, we answer the following questions: 
(1) How do spatial and temporal scales as well as temporal resolution influence prediction accuracy? 
(2) Do reported fishing duration and catch rates depend on the time of the year?
(3) How useful are the considered auxiliary data for predictions in time and space?

We tested the use of citizen-reported data, as reported by an online platform (see Methods), on angler behavior for temporal and spatial predictions and its connection to auxiliary data. The study provides a first step into high-resolution large-scale predictions of overall angler behavior, such as captured by conventional surveys. 

\section{Materials and Methods} \label{s:methods}
Auxiliary data on environment, socioeconomics, and fisheries management and events were used to predict angler behavior across Canada reported on the online platform Angler's Atlas (collected via the website and the mobile app MyCatch). Spatial predictions were made for up to five consecutive years at water bodies in up to three provinces of Canada. Different temporal resolutions as well as spatial subsets were tested. For each prediciton task, nine machine-learning models (General linear regression, support vector regression, k-nearest neighbors, random forest, gradient-boosted regression trees, neural network, general Bayesian network, naive Bayes and tree-augmented naive Bayes) were trained.

\subsection{\emph{Study area and data}} \label{s:methods.1}
Data from the three Canadian provinces of Ontario, British Columbia, and Alberta were considered over a span of five years (2018 to 2022). See SI Methods for differences in economics and climate along the provinces. Only open-water seasons were assessed, which in this study were defined as 184 days per year between the beginning of May and the end of October.

Water bodies comprised rivers and lakes. If a water body was part of multiple fisheries management zones (established by the provincial governments of Canada, (\cite{Ontario2019, Alberta2020, BC2019})) it was split into different sections and each section was assigned a unique water body ID. In the following, the term ``water body" can encompass both a whole water body or, when specified, a section of a water body.

\subsubsection{\emph{Citizen-reported angler behavior}} \label{s:methods.11}
The online platform Angler's Atlas and its mobile phone app ``MyCatch" recorded almost 27,000 trips in Ontario, about 16,000 trips in British Columbia, and almost 13,000 trips in Alberta between 2018 and 2022.
Only data from completely reported angler trips were considered. A trip was considered to be completely reported if information on location and all target variables was included. As organized fishing events can produce their own set of alternative dynamics, a fishing event initiated by Angler's Atlas took place at a water body, the reported trips were excluded for the time period of the event. Additionally, trips with unrealistic fishing durations exceeding 24 hours in a single day were removed from the data set.




\subsubsection{\emph{Features on the environment, socioeconomics, and fisheries management}} \label{s:methods.12}

The features, defined as the auxiliary data used to predict the target variables, covered three different categories: (i) the \emph{environment}, (ii) the \emph{socioeconomics} in the surrounding area of a water body and (iii) the \emph{fisheries management} of the water body including events. For more details and references that support the inclusion of the features see Tables \ref{tab:Features_Refs}, \ref{tab:Features}.  

\emph{Environmental features} comprised information on the terrain and the weather. The water body type (lake or river) was captured in a binary indicator feature. The surface area and shoreline of lakes were received from the internal Angler's Atlas database. For river sections, the surface area was set to zero because no data on the river width were available, and the shoreline was defined as the length of the section multiplied by two. The elevation of the water body and daily weather was obtained from the simulation model BioSIM (\cite{Regniere2017}). See SI Methods for details. Daily growing degree days were calculated using the equation in \textcite{Uphoff2013} and summing days with an average air temperature $\geq$ 5°C starting on April 1st of the respective year (\cite{Uphoff2013, Chezik2014}). 

\emph{Socioeconomic features} comprised economic and demographic aspects in the surrounding area of the water body as well as information on the infrastructure in terms of accessibility of the water body. Variables also covered information on Covid-19 as it led to changes in angler behavior (\cite{Trudeau2022, Midway2021, Howarth2021, Gundelund2021MP}). See Tables \ref{tab:Features_Refs}, \ref{tab:Features} and SI Methods for details. The temporal and spatial resolutions and extents varied among the variables (Table \ref{tab:Features}). As not all of the applied machine-learning methods could deal with missing values, the value of a variable was set to zero if the information was not available. 

\emph{Features on fisheries management and events} provided insights into the management practices, regulations, and temporal considerations that may impact angler behavior. Water bodies could be closed or partially closed (some sections) for fishing, could have regulations on fish sizes or the number of fish that could be kept (bag limit) and could have the catch-and-release regulation. Data on stocking events of hatchery fish were available for Ontario. For Alberta and British Columbia, daily active fishing licences were available. A zero value of the binary regulation features (``Fish size limit", ``Bag limit", ``Catch-and-release", ``Stocking") had two meanings, namely either no respective regulation at the water body, or missing information. The feature ``Weeks since last stocked" was set to zero when the water body was not stocked at all, indicated in the feature ``Stocking event in the year". 

Note that the location of a water body was not used as a feature in the model, but was necessary for the calculation of some features such as weather and distances. It was defined by the latitude and longitude of the water body's centroid.
 
\subsection{\emph{Model prediction tasks}} \label{s:methods.2}
The study comprised different prediction tasks regarding (1) the target variable, (2) the temporal resolution of the features, (3) the spatial extent of the studied area, and (4) the temporal extent of the considered angler behavior. 

\subsubsection{\emph{Target variables}} \label{s:methods.21}
Two aggregated variables were used for examining reported angler behavior:

(1) The total fishing duration $D_{w,t}$ at water body $w$ and time $t$, representing angler pressure. This measurement was obtained by summing the hours spent fishing by all anglers:
\begin{equation*}
D_{w,t} = \sum_i{D_{i(w,t)}}
\end{equation*}
with $D_i$ being the fishing duration of angler $i$ at the water body $w$ at time $t$.

(2) The mean catch rate, denoted as $C_{w,t}$, which  reflects the angler's success in catching fish. It was calculated as the average number of fish caught per hour:
\begin{equation*}
    C_{w,t} = \frac{1}{i}\sum_i{\frac{catches_{i(w,t)}}{D_{i(w,t)}}}
\end{equation*}
with $i$ being a reported trip that includes information on the number of fish caught $catches_i$ and the fishing duration $D_{i(w,t)}$ of the angler at water body $w$ and time $t$.

Independent models were trained to predict (1) total fishing duration $D_{w,t}$ and (2) mean catch rates $C_{w,t}$.

\begin{figure}[H]
  \includegraphics[width=1\linewidth]{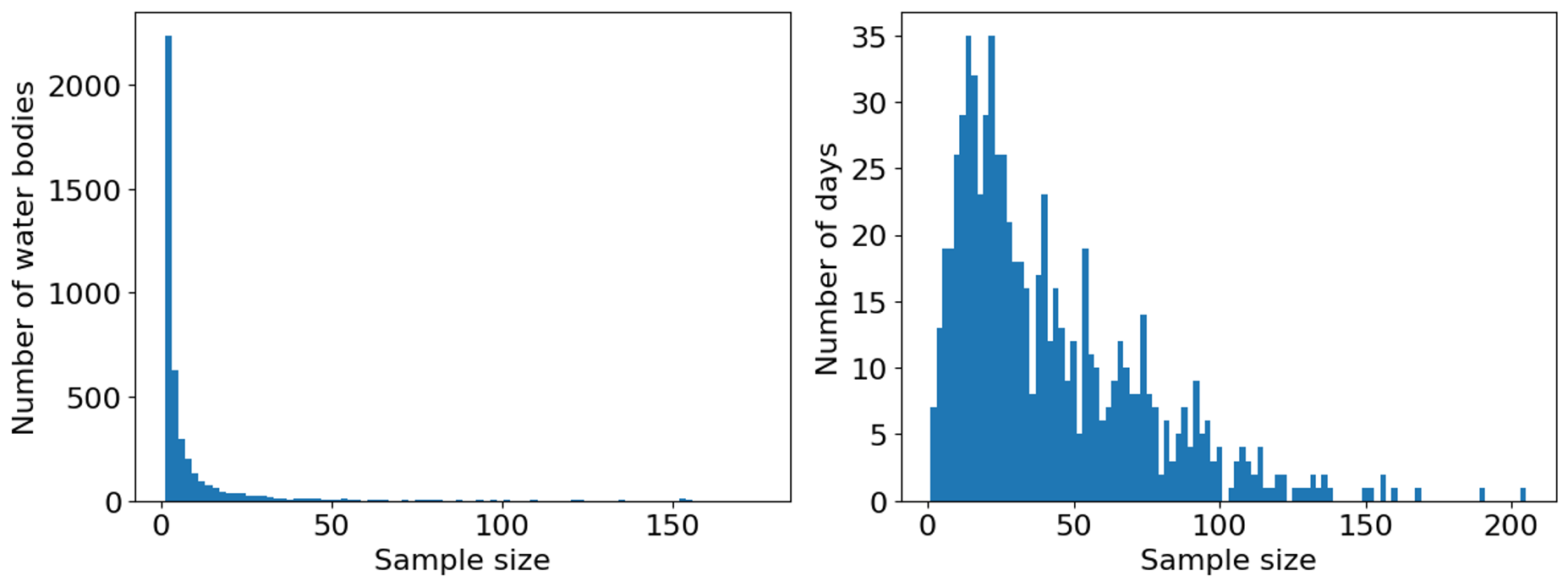}
  \caption{Number of samples per water body (over open-water seasons of five years) and per day (over the three provinces). On average, a water body has 7 samples (days with reports) and a day has 41 samples (water bodies).}
  \label{fig:SampleSize}
\end{figure}

\begin{figure}[H]
  \includegraphics[width=1\linewidth]{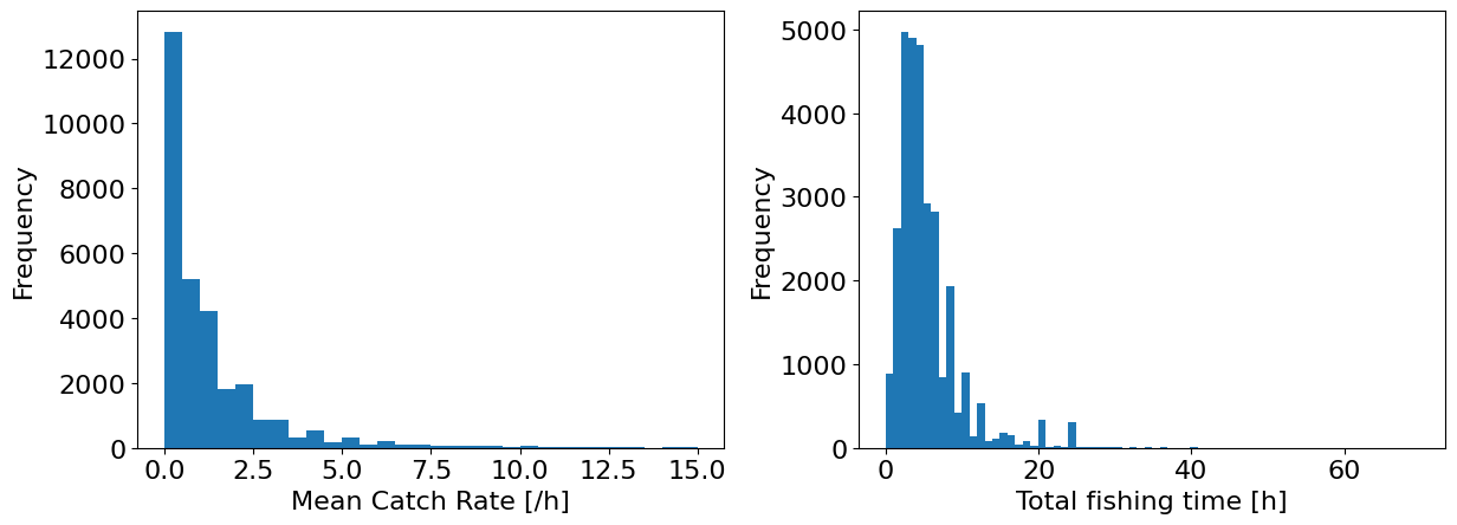}
  \caption{Distribution of the mean catch rate and total fishing duration in samples across the three provinces (entire data set). 
}
  \label{fig:FreqTarVars}
\end{figure}

The data set used for predictions included water bodies and dates with at least one report on Angler's Atlas, which resulted in 30,193 daily aggregated samples over the three provinces (see Table \ref{tab:DatasetSize} for details). On average, 41 samples per day across all water bodies, and seven samples per water body over the entire time span were available, whereby the majority of days and water bodies had fewer samples (Fig. \ref{fig:SampleSize}, Table \ref{tab:DatasetSize}). Both target variables, i.e., total fishing duration and the mean catch rate, were exponentially distributed at a daily resolution (Fig. \ref{fig:FreqTarVars}).

All features were used to predict catch rates and fishing durations as the target variables could influence each other (\cite{Kuparinen2010}).

\begin{table}[]
\caption{Number of water bodies and samples in the data sets used for predictions.}\label{tab:DatasetSize}
\begin{tabular}{|c|r|r|r|r|}
\hline
Subset &
  \multicolumn{1}{c|}{\begin{tabular}[c]{@{}c@{}}\# Water bodies\end{tabular}} &
  \multicolumn{1}{c|}{\begin{tabular}[c]{@{}c@{}}\# Daily \\ samples\end{tabular}} &
  \multicolumn{1}{c|}{\begin{tabular}[c]{@{}c@{}}\# Weekly \\ samples\end{tabular}} &
  \multicolumn{1}{c|}{\begin{tabular}[c]{@{}c@{}}\# Monthly  \\ samples\end{tabular}} \\ \hline
Total  & 4,147   & 30,193   & 21,840   & 15,278  \\ \hline
ON     & 2,202   & 15,920   & 11,529   & 7,933   \\ \hline
AB     & 540     & 5,981    & 3,864    & 2,479   \\ \hline
BC     & 1,405   & 8,292    & 6,447    & 4,866   \\ \hline
Lakes  & 3,346   & 23,404   & 17,749   & 12,692  \\ \hline
Rivers & 801     & 6,789    & 4,091    & 2,586   \\ \hline
\multicolumn{1}{|c|}{\begin{tabular}[c]{@{}c@{}}Region with     \\ most samples\end{tabular}} & 623  & 3,789  & 2,943  & 2,110 \\ \hline
\multicolumn{1}{|c|}{\begin{tabular}[c]{@{}c@{}}Water body with \\ most samples\end{tabular}} & 1     & 186   & 72     & 24    \\ \hline
\end{tabular}
\end{table}

\subsubsection{\emph{Different temporal resolutions}} \label{s:methods.22}
\begin{figure}[H]
  \includegraphics[width=1\linewidth]{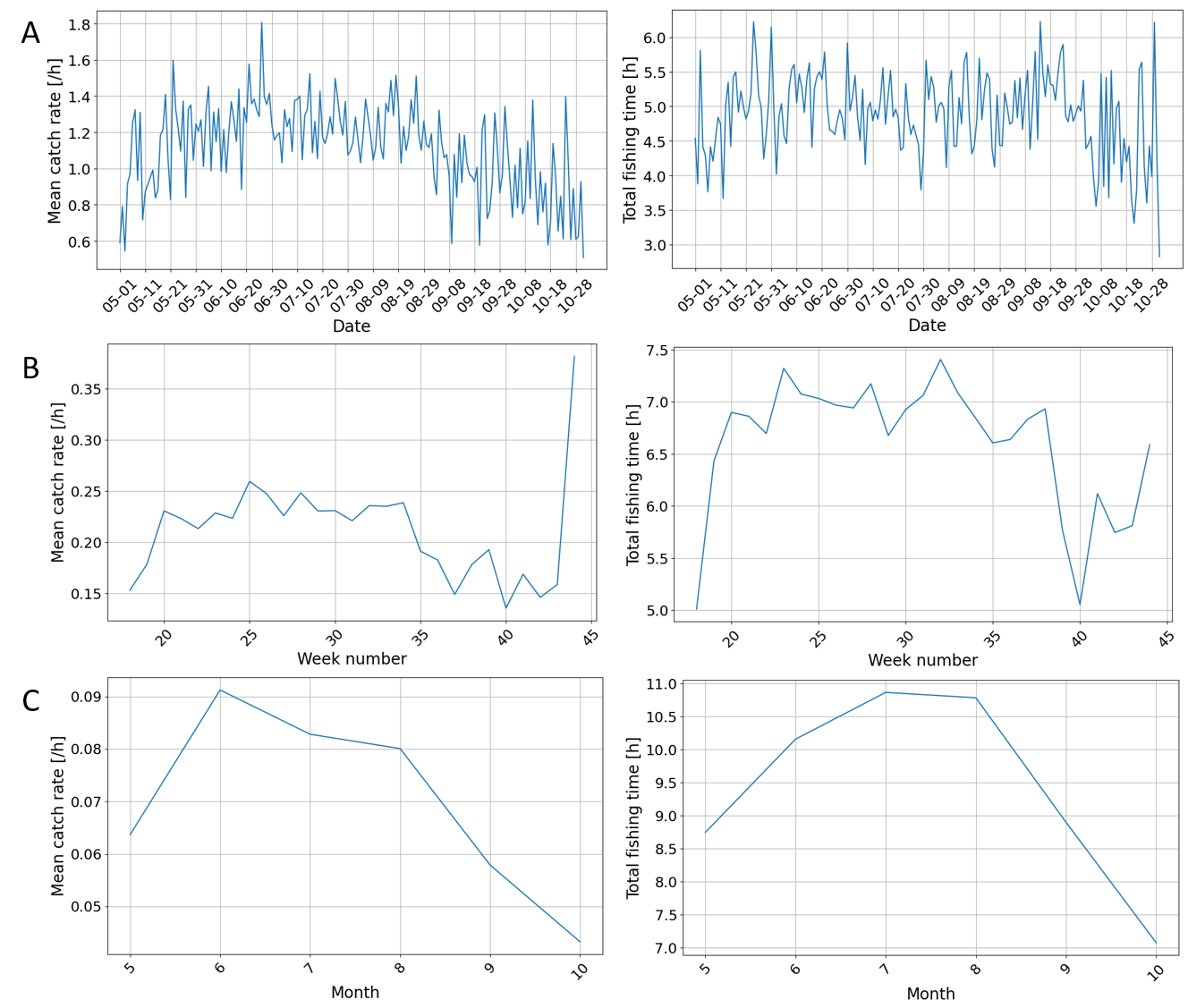}
  \caption{Inter-annual curve of the mean catch rate (left) and total fishing duration (right) in samples across the three provinces (entire data set). The line shows the mean at the specific (A) date, (B) week number and (C) month over 5 years (2018-2022) and all water bodies.
}
  \label{fig:AnnCurveVars}
\end{figure}

Daily, weekly and monthly features and target variables were tested. Daily features were averaged or summed up to receive weekly and monthly resolutions (Table \ref{tab:Features}, Fig. \ref{fig:AnnCurveVars}). Consequently, the sample sizes of the data sets decreased (Table \ref{tab:DatasetSize}).

\subsubsection{\emph{Different spatial extents}} \label{s:methods.23}
Models were trained and tested on different spatial extents (Table \ref{tab:DatasetSize}). Data sets comprised (1) single water bodies (20 subsets of water bodies with largest sample sizes), (2) single regions (20 subsets with regions of largest sample sizes), (3) single provinces (three subsets), (4) only lakes or only rivers across the three provinces (two subsets) and (5) the total dataset (including all water bodies across the provinces British Columbia, Alberta and Ontario).
Models on a single water body made only temporal predictions based on temporal and spatio-temporal features.
Regions corresponded to the fisheries management zones established by the provincial government (19 regions in Ontario (\cite{Ontario2019}), 9 regions in British Columbia (\cite{BC2019}) and 11 regions in Alberta (\cite{Alberta2020})).
If a feature had an equal value for all samples in a subset it was removed from the model (before or after discretization; e.g., the whirling disease was only present at some water bodies in Alberta).

\subsubsection{\emph{Different temporal extents}} \label{s:methods.24}
Different temporal extents were considered in the models (Table \ref{tab:DatasetSize}). Besides models considering (1) the entire five years, data subsets of (2) only a month, (3) only one week and (4) only one day in a specific spatial extent were trained and tested. Models using subsets (2)-(4) made only spatial predictions based on spatial and spatio-temporal features, whereby the temporal aggregated features on the respective temporal extent were used. For (2)-(4), 20 subsets of time stamps with largest sample sizes were used, respectively.

\subsection{\emph{Model training and testing}} \label{s:methods.3}

For model training and testing, 3-fold crossvalidation was applied by splitting the data sets into three parts for training and testing. 

If the data set consisted of only one water body, the split was done temporally by three time periods with equal sample sizes to account for potential temporal autocorrelation. Two time periods were used for model training, and the remaining time period for model testing.

If the data set consisted of a single time step (either a day, a week or a month), the spatial split was done randomly in ratio 2:1. No spatial autocorrelation between the water bodies was assumed.

If the data set consisted of more than one water body and the entire time span, the split was done spatially based on water bodies ``batches", with each batch containing the reported data of a single water body for the entire five years. The data was then split randomly by water bodies, whereby two-thirds of the water bodies were included in the training set and the remaining third in the test set. 

For each of the prediction tasks, nine independent machine-learning methods were trained and tested: six methods for regression and three methods for classification. 
For the regression methods, continuous features were standardized based on the respective training data set by removing the mean and dividing by the standard deviation (StandardScaler or z-score). 
For the classification methods, the features and the target variable were discretized into two bins whereby each bin had the same width in the span of possible values for the feature. See SI Methods for details on the machine-learning methods used for regression and classification.

To compare the prediction performance of the regression models with the classification models, the predicted target values were discretized into two bins based on the median value of the samples in the corresponding training set. 

The accuracy score of the test set $i$ was calculated for the evaluation of each prediction model:
\begin{equation*}
    \text{Accuracy}_{\text{model}_i} = \frac{\text{Number of correct predictions}}{\text{Number of total predictions}}.
\end{equation*}

\subsection{Evaluation of predictions} \label{s:methods.4}

For each prediction task (specific target variable, temporal resolution, spatial extent and temporal extent) and machine-learning method $j$, the mean over the three models using the different training-test splits was computed to receive a mean accuracy score:
\begin{equation*}
    \text{Accuracy}_{\text{method}_j} = \frac{1}{3} \sum_{i=1}^3 \text{Accuracy}_{\text{model}_i}.
\end{equation*}

Of these mean accuracy scores, the best-performing machine-learning method was determined and chosen to compare the performance of the different prediction tasks:

\begin{equation*}
    \text{Accuracy}_{\text{task}} = \max_{j \in \{1, 2, \ldots, 9\}} \text{Accuracy}_{\text{method}_j}.
\end{equation*}

\subsection{\emph{Modified prediction tasks}} \label{s:methods.5}
To analyze the role of inter-annual variation in fishing behavior and the importance of the considered features, modified prediction tasks were tested: (i) predicting with additional different temporal features to improve the prediction performance of the models and (ii) predicting without any features.
The predictions were evaluated and compared as described in Methods \ref{s:methods.3}.

\subsubsection{\emph{Predictions with additional features}} \label{s:methods.52}
Additional temporal features were added to the data set: 
(i) A data set with the additional feature Julian day (ranging from 121 to 305), 
(ii) a data set with the additional feature week number (ranging from 17 to 44), 
(iii) a data set with the additional feature month (ranging from 5 to 10), and
(iv) a data set with the additional feature of the number of water body website views on the online platform Angler's Atlas on the previous seven days (received from Google Analytics). Seasonal variation and temporal autocorrelations in fishing behavior and catch rates have been detected in previous studies (\cite{Heermann2013, Trudeau2021, McCormick2023}).

The data sets (i) to (iii) included more information on the date of the samples, and the data sets (iv) also included information on the angler behavior itself in the previous time.

\subsubsection{\emph{Predictions without features}} \label{s:methods.54} 
In this approach, the values of the target variable were predicted without using auxiliary features. Instead, the predicted value $x_{w,t}$ for each sample (water body $w$ at time step $t$) in the test set corresponded to the mean of the values of all samples (water bodies $s$) in the same time step $t$ in the training set: 
\begin{equation*}
x_{w,t}=\frac{1}{n}\sum{x_{s,t}, s=1,...,n}
\end{equation*}
where $n$ represents the number of samples (water bodies) in the training set. Hence, the predicted values of the target variable in the test set were assumed to be equal across all water bodies and varied only in time. If a time step in the test set was not available in the training set, the predicted value of the target variable for this time step was set to 0.

By comparing the performance of this ``null model" to the other approaches, the use of the auxiliary variables in terms of the contribution of valuable information to the predictions could be assessed.

\section{Results} \label{s:results}

\begin{figure}[H]
  \includegraphics[width=0.9\linewidth]{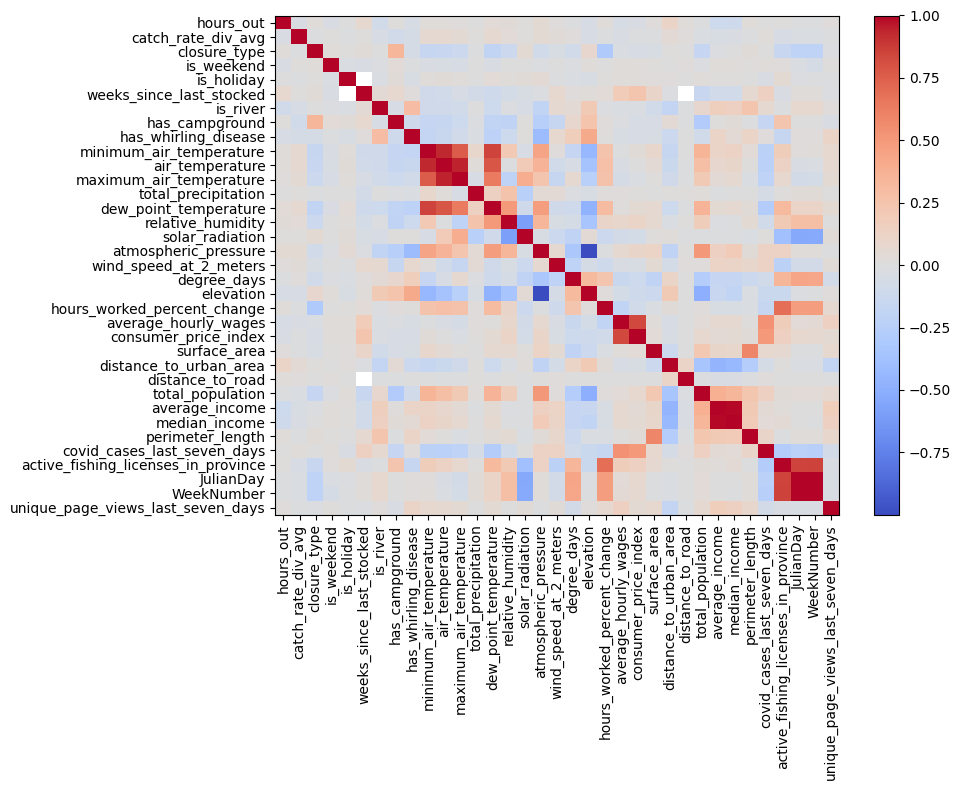}
  \caption{Pearson correlation coefficients between the considered features and target variables. Features were not discretized or standardized. Samples from all three provinces were considered.
}
  \label{fig:PearsonCorr}
\end{figure}

Machine learning models were trained to predict low or high mean catch rates and short or long total fishing durations. The prediction performances of the target variables respond similarly to changes in the spatial or temporal extent, or temporal resolution although they were not correlated (Fig. \ref{fig:PearsonCorr}).

\subsection{Temporal and spatiotemporal predictions}

\begin{figure}
  \includegraphics[width=1\linewidth]{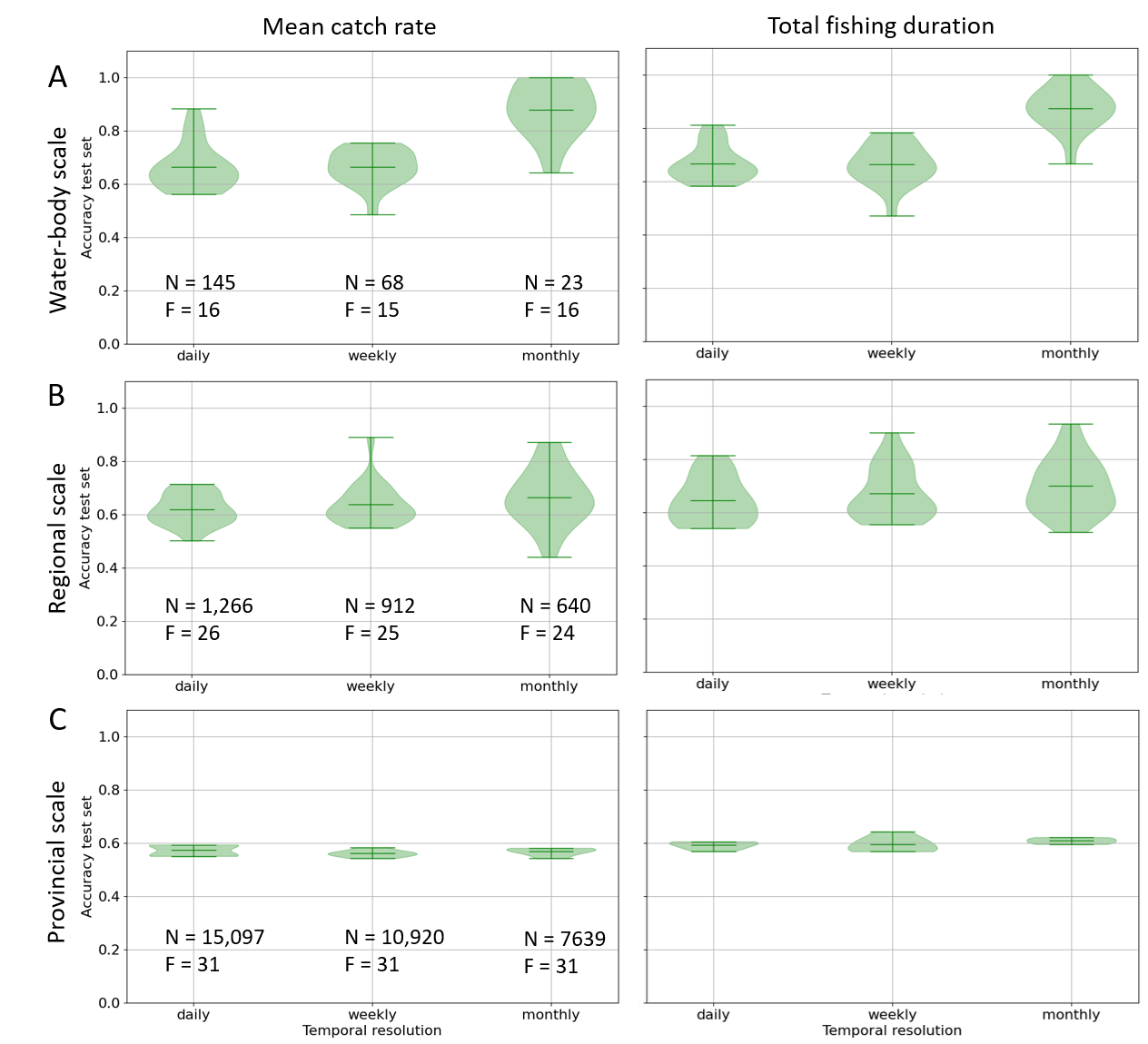}
  \caption{Prediction performance over time (and space) for the mean catch rate (left) and total fishing durations (right). Accuracy scores on test sets of data subsets for (A) the 20 water bodies with the most samples, (B) the 20 regions with the most samples, and (C) the three considered provinces (AB, BC and ON), water body type subsets (lakes, rivers) and the entire data set. Colors show different temporal aggregation levels of reported fishing trips. Marks show minimum, maximum and mean accuracy scores. N: Mean number of data samples for model training and testing, F: Mean number of features in the models. See Fig. \ref{fig:BasicAnalysis_TempRes} for more details.}
  \label{fig:BasicAnalysis_TempRes_violin}
\end{figure}

\begin{figure}[H]
  \includegraphics[width=1\linewidth]{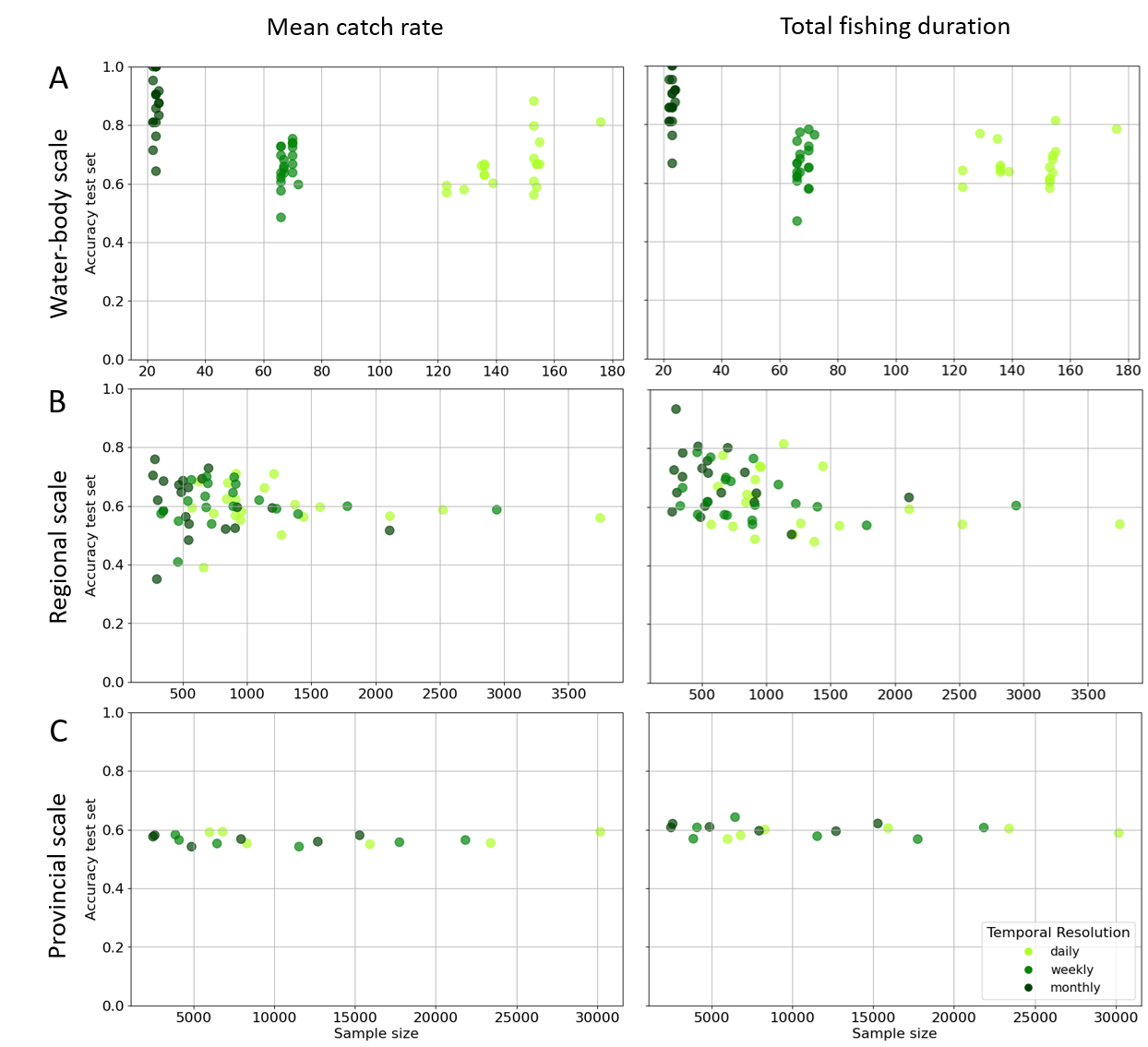}
  \caption{Prediction performance over time (and space) for the mean catch rate (left) and total fishing durations (right). Accuracy scores on test sets of data subsets for (A) the 20 water bodies with the most samples, (B) the 20 regions with the most samples, and (C) the three considered provinces (AB, BC and ON), water body type subsets (lakes, rivers) and the entire data set. Colors show different temporal aggregation levels of reported fishing trips.}
  \label{fig:BasicAnalysis_TempRes}
\end{figure}

Prediction performance varied between the different considered temporal resolutions of the samples and the spatial extents (Figs. \ref{fig:BasicAnalysis_TempRes_violin}, \ref{fig:BasicAnalysis_TempRes}). On average, models on the province level used 11218 samples and 31 features, on the regional level 939 samples and 25 features, and on the water-body level 79 samples and 16 features. Note that models on the water-body level only predicted in time, whereas models using water bodies of regions or higher extents did predict in time and space.

Including water bodies from an entire province, all three provinces, or only a certain water body type (lakes or rivers) across the three provinces in the model revealed no improvement in prediction performance in comparison to the models on a regional level (Figs. \ref{fig:BasicAnalysis_TempRes_violin}C, \ref{fig:BasicAnalysis_TempRes}C). The average accuracy score over all temporal resolutions and data sets for the mean catch rate was 57\%. The maximum accuracy score was 59\%, from a model with daily values considering only rivers. 

Regional predictions of mean catch rates of all water bodies resulted in prediction performances close to 60\% independent of the temporal resolution (Figs. \ref{fig:BasicAnalysis_TempRes_violin}B, \ref{fig:BasicAnalysis_TempRes}B). 

On the water-body level, predicting monthly mean catch rates resulted in the best accuracy score of up to 100\% (88\% on average) with on average 23 monthly samples (Figs. \ref{fig:BasicAnalysis_TempRes_violin}A, \ref{fig:BasicAnalysis_TempRes}A left). Using weekly samples caused a drop in accuracy to a maximum of 75\% (66\% on average). Daily predictions showed with an accuracy score of up to 88\% (66\% on average) along the considered water bodies similar performance as in the weekly resolution.

Model predictions of total fishing durations had a similar performance (Figs. \ref{fig:BasicAnalysis_TempRes_violin} right, \ref{fig:BasicAnalysis_TempRes} right).

\subsection{Spatial predictions} \label{s:results.1}

\begin{figure}
  \includegraphics[width=1\linewidth]{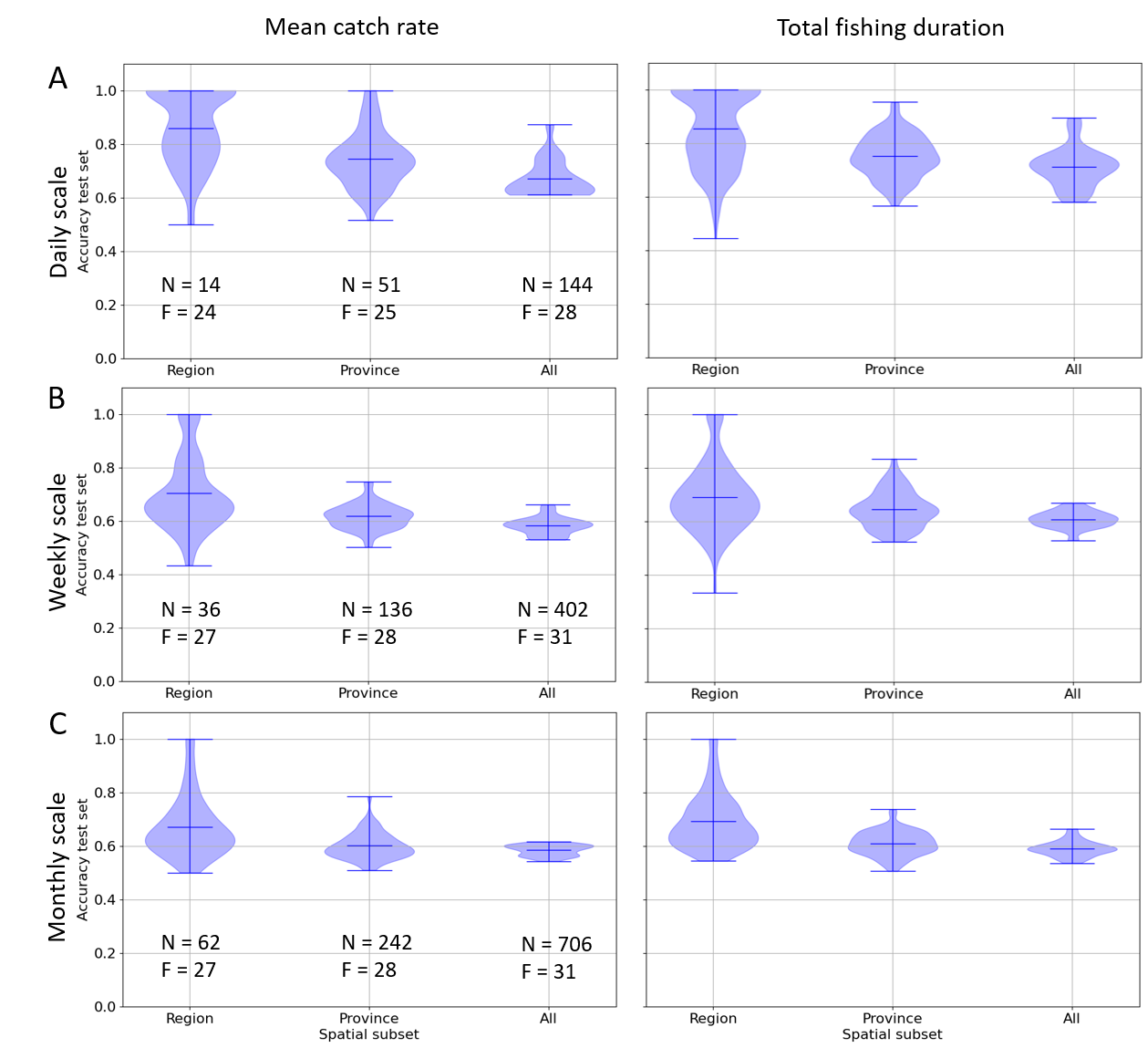}
  \caption{Prediction performance over space for the mean catch rate (left) and total fishing duration (right). Accuracy scores on test sets of data subsets for (A) the 20 days, (B) the 20 weeks, and (C) the 20 months with most samples in a specific area. Bodies show different spatial extents of reported fishing trips (single regions, single provinces and entire study area (three provinces). Marks show minimum, maximum and mean accuracy scores. N: Mean number of data samples for model training and testing, F: Mean number of features in the models. See Fig. \ref{fig:BasicAnalysis_SpatRes} for more details.}
  \label{fig:BasicAnalysis_SpatRes_violin}
\end{figure}

\begin{figure}[H]
  \includegraphics[width=1\linewidth]{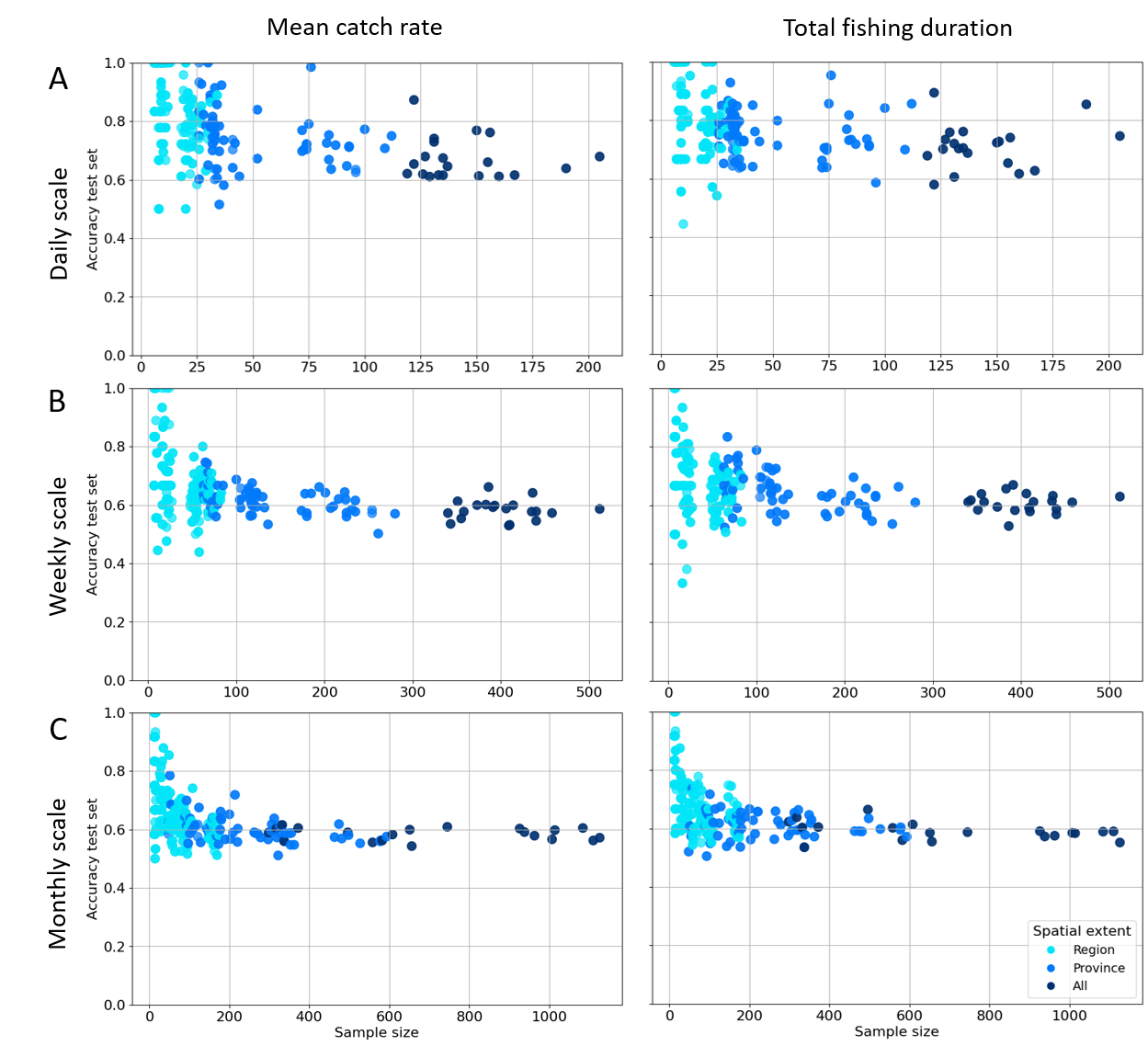}
  \caption{Prediction performance over space for the mean catch rate (left) and total fishing duration (right). Accuracy scores on test sets of data subsets for (A) the 20 days, (B) the 20 weeks, and (C) the 20 months with most samples in a specific area. Colors show different spatial extents of reported fishing trips (single regions, single provinces and entire study area (three provinces).}
  \label{fig:BasicAnalysis_SpatRes}
\end{figure}

Considering a single time step (either a day, a week or a month) in the prediction models resulted in better prediction performance compared to spatiotemporal predictions. The sample sizes, i.e., the number of water bodies, used for model predictions decreased with finer temporal scale and smaller spatial extent (Figs. \ref{fig:BasicAnalysis_SpatRes_violin}, \ref{fig:BasicAnalysis_SpatRes}). Mean accuracy scores slightly increased with a finer temporal resolution. For instance, for mean catch rates the mean accuracy score was 80\% on a single day, 67\% in a single week, and 64\% in a single month with samples in respective resolution. Within a temporal resolution, smaller spatial extents led to better predictions. For instance, for daily mean catch rates the mean accuracy score was 86\% for regions, 75\% for provinces, and 67\% for all three provinces (Figs. \ref{fig:BasicAnalysis_SpatRes_violin}A, \ref{fig:BasicAnalysis_SpatRes}A).

\subsection{Predictions with additional features}
The inclusion of additional features such as the Julian day, the week number, the month, and the number of water-body website views, did not remarkably change the prediction performance (Table \ref{tab:AdditionalFeatures}). Along all spatiotemporal prediction tasks, prediction accuracy increased in maximum by 1.6\% by adding the month to water-body predictions of total fishing duration, and mostly decreased, in maximum by 5.5\%.

\subsection{Predictions without features} 
\begin{figure}[H]
  \includegraphics[width=0.9\linewidth]{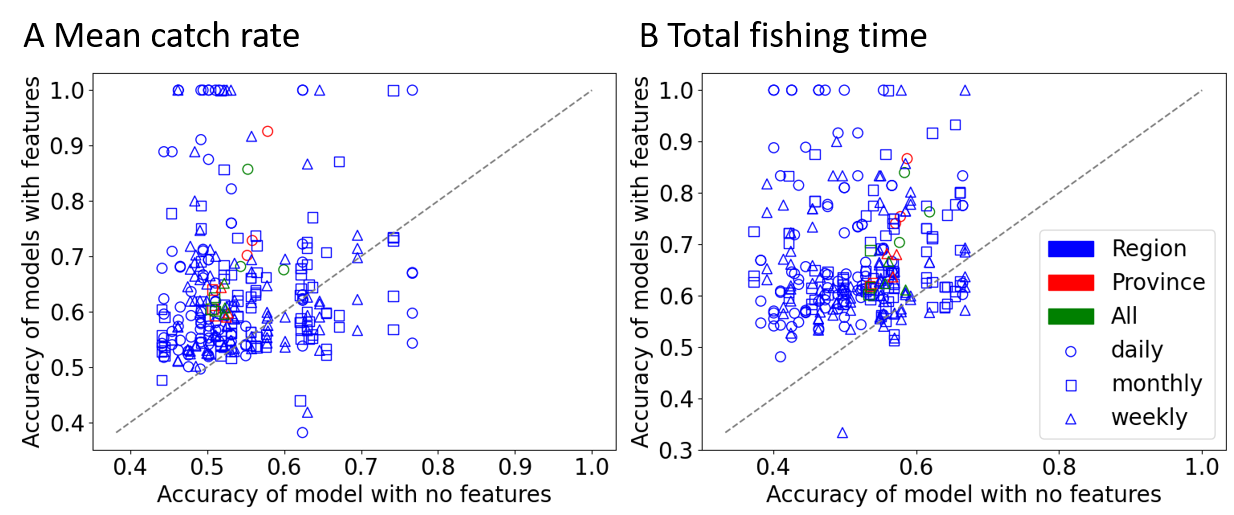}
  \caption{Comparison of accuracy scores between the models using features and the temporal mean of the target variable. Scatter plots contain (A) ``Mean catch rate" and (B) ``Total fishing duration" prediction scores. The green symbols show results for all three provinces, and subsets of only rivers and only lakes across all three provinces.
}
  \label{fig:TargetbyTarget}
\end{figure}

The simple approach of predicting the mean catch rates or total fishing durations by averaging their values in the training set in a specific time step resulted in worse performance for most spatial subsets and temporal resolutions (Fig. \ref{fig:TargetbyTarget}). Only in a few regions the predictions of the simple approach showed equal or better accuracy scores.

\section{Discussion} \label{s:discussion}

For the prediction of angler behavior in terms of mean catch rates and total fishing durations reported on an online platform, several spatial and temporal scales as well as temporal resolutions and multiple machine-learning methods were tested to find spatiotemporal relationships between citizen-reported angler behavior and environmental, socioeconomic, fisheries management, or event features. Overall, maximum prediction performances were achieved for non-spatially monthly predictions at single water bodies with a mean accuracy score of 88\% for the mean catch rate and 87\% for the total fishing duration.  
On a weekly or daily resolution and provincial or regional scale, the machine-learning models gained only a low predicition accuracy of about 60\%. Spatial predictions were most accurate on a daily regional scale. 
Adding temporal features related to the day or season of the year and angler behavior in the previous week had a negligible impact on prediction performance. 
Comparing the model predictions with the simple spatial averages at a given time step of the angler behavior variable showed that the considered features contributed useful information to the predictions.

The similar predictive performance between mean catch rates and total fishing durations indicate a strong impact of the sample size and the selected features. 
The accuracy score of angler behavior predictions increased with smaller spatial extent and broader temporal resolution. With this, the data sample sizes decreased which could lead to overfitting given the large number of features of the models. Some prediction models on a monthly resolution or that considered only single water bodies had more features than data samples. Lower sample sizes can restrict the variability in the test set when validating the model. For large spatial extents and fine temporal resolutions with higher variability, such as province-wide predictions on a daily resolution, the models were likely underfitted due to insufficient data for model training. To enhance the analysis, increased promotion and greater involvement of anglers on the online platform and mobile application can lead to a more comprehensive data set, allowing the machine-learning methods to better uncover relationships and make predictions (\cite{Skov2021}).
Moreover, although mean catch rates and total fishing durations were not directly correlated, there are potential indirect relationships across the features considered. Feature selection can be applied to remove irrelevant and redundant data to improve model performance of some of the machine-learning methods (\cite{Jie2018}).


Limitations in the used features can also be a reason for the poor prediction performance. The features were only available in different spatial and temporal resolutions, ranging from province-wide, annual values to water-body-based, daily resolution. A field study revealed that even within a single water body, the heterogeneity of environmental conditions can significantly influence angler behavior (\cite{Dippold2020}).
Furthermore, the selected features primarily focused on environmental and demographic impacts but overlooked information such as present fish species, fish abundances, and environmental parameters influencing fish abundance (e.g., dissolved oxygen concentration, prey availability, and water clarity) which can also affect anglers' choices of fishing sites (\cite{Dippold2020, Hunt2019Elsevier, Hunt2019RFSA, Heermann2013}). Besides this, popular fishing methods, such as trolling or casting, fishing experience and skills, or gear restrictions at a water body can also play a role (\cite{Bade2022, Dippold2020, Dabrowksa2017}). Social media activity, fishing trends, previous anglers satisfaction, and events such as tournaments, which are also related to catch expectations, can be relevant for the prediction of angler behavior (\cite{Matsumura2019, Dabrowska2014, Powers2016}). However, the inclusion of such features can be challenging due to cost limitations, missing information, or the absence of appropriate metrics, for example, for data related to social media.

A low impact of temporal variables on prediction performance is in contrast to previous studies showing that the season had an impact on catch rates and angler effort during the open-water season (\cite{Heermann2013, Trudeau2021}). The seasonal input and temporal autocorrelation might already be covered in other features, such as air temperature or sold fishing licences in a province, so that adding the temporal variables resulted in no more additional information for the models (\cite{McCormick2023}).

The study builds a first attempt to predict angler behavior on a large scale using easily accessible auxiliary data. The analysis was restricted to the prediction of citizen-reported data that was gathered with a uniform methodology across a large spatial and temporal scale. Still, the citizen-reported data were biased towards a certain fraction of anglers that used the online platform and mobile application and were willing to share their fishing activity. The fact that the models were unable to predict these data at a larger spatial and finer temporal scale underscores the important new information they could provide about fishing behavior that was uncaptured by the auxiliary data. 

Data on general angler behavior that is not biased to anglers willing to share their trips on online platforms can be gathered through conventional methods like creel surveys or camera monitoring (\cite{Askey2018, Hunt2019Elsevier, Morrow2022}).
In a following study, the presented methods can be used to predict the conventionally gathered angler behavior, using the citizen-reported angler behavior from the online platform as input features for the models (\cite{Trudeau2021}). Such models could be used for temporal predictions into the future and for spatial predictions at water bodies with no or only a limited amount of conventionally gathered data.

\section{Conclusion}\label{s:conclusion}
In summary, the study revealed the use of auxiliary data on the environment, socioeconomics, and fisheries management and events in combination with advanced machine-learning methods to receive valuable predictions of aggregated citizen-reported angler behavior at freshwater bodies across Canada. The study builds a first attempt to predict angler behavior on a large scale and can be expanded into different directions, such as the prediction of non-biased angler behavior received through conventional surveys or the prediction of angler preferences based on socioeconomic angler backgrounds.

\section*{Acknowledgements}
We acknowledge the support of the Government of Canada’s New Frontiers in Research Fund (NFRF), NFRFR-2021-00265. MAL gratefully acknowledges an NSERC Discovery Grant.

We acknowledge Joel Knudsen, Jamie Svendsen, Azar T. Tayebi and Clayton Green for their support on gathering and preparing the data. 

We acknowledge Rob Corrigan from the Government of Alberta, and Adrian Clarke and Adeleida Bingham from the Freshwater Fisheries Society of BC for providing numbers of fishing licence sales.

We acknowledge members of the Lewis lab for helpful feedback and suggestions.


\newpage
\beginsupplement

\if0\blind
{
    \title{\bf Supplementary Information: Can machine learning predict citizen-reported angler behavior?}
    
    \author{Julia S. Schmid$^1$, Sean Simmons$^2$, Mark A. Lewis$^{1,3,4,5}$, Mark S. Poesch$^5$, Pouria Ramazi$^6$ \\
    \small
        $^1$Department of Mathematical and Statistical Sciences, University of Alberta, Edmonton, Alberta, Canada \\
        \small
        $^2$Angler’s Atlas, Goldstream Publishing, Prince George, British Columbia, Canada \\
        \small
        $^3$Department of Mathematics and Statistics, University of Victoria, Victoria, British Columbia, Canada \\
        \small
        $^4$Department of Biology, University of Victoria, Victoria, British Columbia, Canada \\
        \small
        $^5$Department of Biological Sciences, University of Alberta, Edmonton, Alberta, Canada \\
        \small
        $^6$Department of Mathematics and Statistics, Brock University, St. Catharines, Ontario, Canada}
    \maketitle
} \fi

\newpage
\section{SI Methods}

\subsection{Comparison of provinces}
The study area comprised three Canadian provinces: Ontario, British Columbia, and Alberta. Ontario is the largest province in terms of area, covering 1,076,395 km2, and has the highest population density of 15.94/km2. British Columbia has a slightly smaller total area of 944,735 km2, with a notably lower population density of 5.41/km2. Alberta, with a total area of 661,848 km2, has a higher population density of 6.66/km2 compared to British Columbia.

The climate in Ontario is humid continental with hot, humid summers and cold winters, British Columbia showcases a diverse climate ranging from coastal mildness to interior continental conditions, and Alberta has a continental climate with cold winters and warm summers.

\subsection{Features on weather from BioSIM}
We used the software tool BioSIM 11 to receive daily weather data and the elevation of each water body (\cite{Regniere2017}, Table \ref{tab:Features}). BioSIM selected the four nearest weather stations for each water body (based on the centroid of the water body) for interpolations and adjusted weather data for differences in elevation, latitude and longitude. Historical daily weather observations were used (Open Topo Data API Nasa srtm 30 m) and the bi-linear interpolation method was applied in the observation-based Climatic Daily model. 

\subsection{Distance of a water body to the next urban area and to a road}
City boundaries and roadways were taken from Statistics Cananada (Spatial information products: Boundary files, 2021 and Road network files, 2022). 
The geometry of water bodies were simplified by 0.0005 degrees using the Douglas-Peucker algorithm implemented in ''ST\textunderscore Simplify()" in PostGIS (\url{https://postgis.net}). 
The minimal Cartesian distance between the nearest points for each simplified water body and road or the centroid of a city based on their coordinates was determined using (''ST\textunderscore Distance" in PostGIS). 

\subsection{Surrounding area of a water body for demographic data}
Demographic data was calculated for each water body by considering cities in the surrounding area at different distances. 
A weighted mean of the human population size, and mean and median income in the surrounding area of a water body was computed by considering three different distances (0.6 * 11 km distance + 0.3 * 111 km distance + 0.1 * 555 km distance).

\subsection{\emph{ML methods for regression}} \label{s:si_methods.31}

\begin{figure}[H]
  \includegraphics[width=0.75\linewidth]{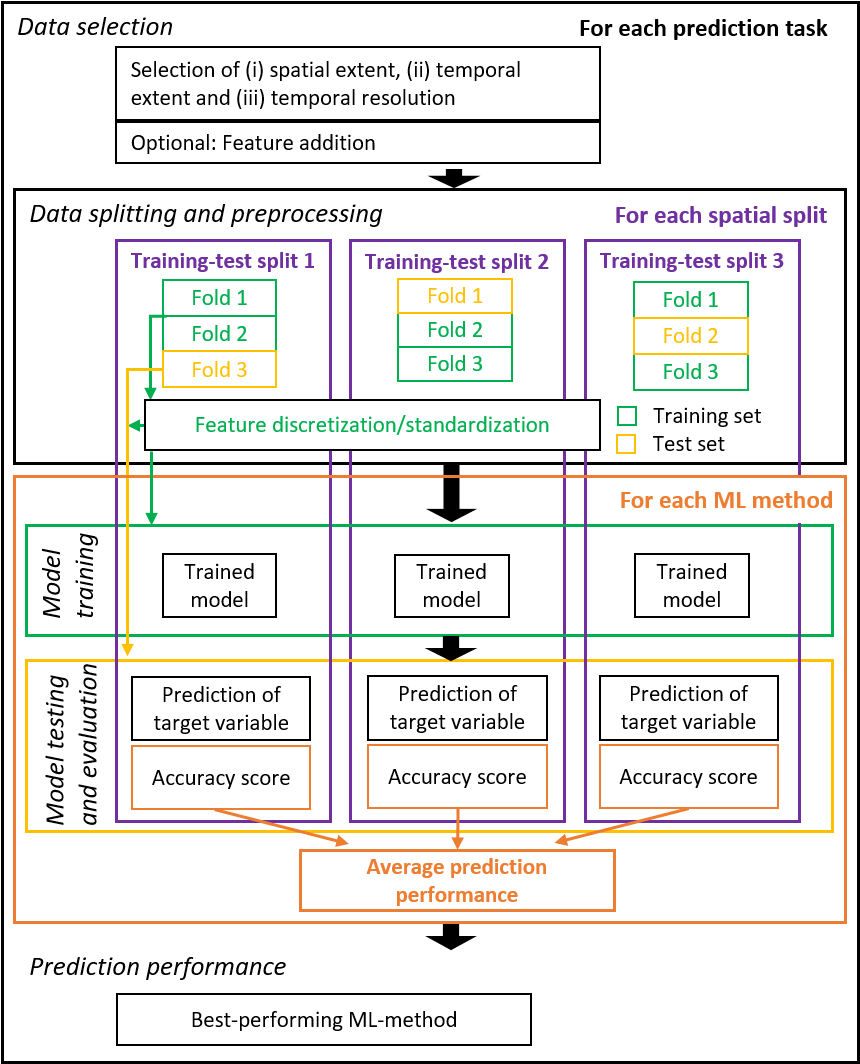}
  \caption{Workflow for training the machine-learning (ML) models and selecting the best-performing 
ML method for a target variable (mean catch rate or total fishing duration) and prediction task.  
}
  \label{fig:Workflow}
\end{figure}

The following six ML methods for regression were applied: (i) a multiple linear regression (MLR), (ii) a support vector regression (SVR) using the radial basis function, (iii) the k-nearest neighbors method (KNN), (iv) a random forest (RF), (v) gradient boosted regression trees (GBRT) and (vi) a neural network (NN). All regression methods used the same set of features to predict the target variables and were trained on the same training sets, respectively (three different training sets for each spatial scale and temporal resolution, see Fig. \ref{fig:Workflow}). 
Analyses were conducted in Python using the library scikit-learn (\cite{Pedregosa2011}). The hyperparameters of the different methods correspond to the default values in the Python library. 

\subsection{\emph{ML methods for classification}} \label{s:si_methods.32}
Bayesian networks (BNs) were applied for predictions of classified target variables (\cite{Ramazi2021MEE}). The learning process of BNs involved two main steps: structure learning and parameter learning. All features were discretized into two bins, based on their mean values in the training set. Similarly, the target variable was discretized into two bins of equal size, using the median value from the training set. 
Functions of the Python library scikit-learn (version 1.1.3) were used. 

Three different BN classifiers according to their strategies to learn the network structure were applied: (i) General BNs learned by the heuristic search approach local hill climbing capturing dependencies between features and the target variable, whereby scores corresponded to the Bayesian Information Criterion (BIC); (ii) naïve Bayes assuming conditional independence between all pairs of features and, hence, covering only dependencies from the target variable to the respective feature; and (iii) Tree-Augmented naive Bayes (TAN) using a tree structure which allows each feature to depend on one other feature besides the target variable (\cite{Friedman1997}).

Two different methods for parameter learning, i.e., the learning of the conditional probability tables, were tested: the (i) Maximum likelihood estimate and (ii) the Bayesian estimation method. 

In total, six BNs were tested (three different structures and two different methods for parameter learning).
BN learning was implemented in Python using the library bnlearn (\cite{Taskesen2020}).

\newpage
\section{SI Figures}

\newpage

\newpage

\newpage

\newpage

\newpage

\newpage

\newpage

\newpage
\section{SI Tables}

\begin{longtable}{|p{6cm}|p{5.0cm}|p{5.0cm}|}
\caption{\label{tab:Features_Refs} Studies that found relationships between the features used for predicting the target variables.}\\
\hline
\textbf{Feature} & \textbf{Catch rate} & \textbf{Fishing duration} \\ \hline
\endfirsthead
\multicolumn{3}{c}{{\tablename\ \thetable{} -- continued from previous page}} \\
\hline
\textbf{Feature} & \textbf{Impact on catch rate} & \textbf{Impact on fishing duration}  \\ \hline
\endhead
\hline \multicolumn{3}{r}{{Continued on next page}} \\ \hline
\endfoot
\hline
\endlastfoot
Environment & & \\
\hline
Minimum air temperature [°C] & \cite{Kuparinen2010} & \cite{Malvestuto1979, Midway2023}\\
Mean air temperature [°C] & \cite{Bade2022, Kuparinen2010, Beamesderfer1995, Shaw2021} & \cite{Malvestuto1979, Midway2023} \\
Maximum air temperature [°C] & \cite{Kuparinen2010} & \cite{Malvestuto1979, Midway2023} \\
Total precipitation [mm] & & \cite{Malvestuto1979, Arlinghaus2017Elsevier}\\
Dew point temperature [°C] &  & \\
Relative humidity [\%] & \cite{Arlinghaus2017Elsevier} &  \\
Solar radiation [watt/m2] & \cite{Shaw2021, Arlinghaus2017Elsevier} & \\
Atmospheric pressure [hPa] &  & \cite{Jensen2022, Arlinghaus2017Elsevier} \\
Wind speed at 2 m [km/h] & \cite{Kuparinen2010, Ho2021, Arlinghaus2017Elsevier} & \cite{Zischke2012} \\
Degree days [°C] & \cite{Dippold2020, Beamesderfer1995} &  \\
River or lake & \cite{Moyle1996} &  \\
Elevation [m] & \cite{Beamesderfer1995, Moyle1996} &  \\
Surface area [m2] & \cite{Kane2023} & \cite{Hunt2007, Hunt2011, Kane2023} \\
Shoreline [m] & \cite{Kane2023} & \cite{Kane2023} \\
\hline
Socioeconomics & & \\
\hline
Human population size in surrounding area [people] & & \cite{Hunt2011} \\
Mean income in surrounding area [CA\$] & \cite{Huppert1989, Capkin2023, Goldsmith2018} & \\
Median income in surrounding area [CA\$] & \cite{Huppert1989, Capkin2023, Goldsmith2018} & \\
Distance to next urban area [m] & \cite{Post2002, Post2008} & \cite{Hunt2007, Hunt2011, Dabrowksa2017, Post2002, Post2008}  \\
Distance to road [m] & & \cite{Hunt2011} \\
Campground availability & & \cite{Hunt2011} \\
Change in work hours due to Covid-19 [\%] & \cite{Huppert1989} & \cite{Midway2023} \\
Average hourly wages & \cite{Huppert1989} &  \\
Consumer price index & \cite{Huppert1989, Parkinson2018} &  \\
Covid cases in the last seven days & & \cite{Midway2023} \\
\hline
Fisheries management and events & & \\
\hline
Bag limitations & & \cite{Goldsmith2018} \\
Fish size limitations & & \cite{Goldsmith2018, Parkinson2018} \\
Catch-and-release regulation & & \cite{Murphy2019, Goldsmith2018, Parkinson2018}\\
Water body closure & & \\
Weekend day or weekday & - & \cite{McCormick2023, Jensen2022, Bucher2006, Parkinson2018} \\
Public holiday (+ connected weekend) & & \cite{Bucher2006} \\
Number of active fishing licenses in the province & & \cite{Hunt2007, Hunt2011} \\
Stocking event in the year & & \cite{Parkinson2018}\\ 
Weeks since the last stocking event [weeks] & & \cite{Parkinson2018} \\
Whirling disease detected & & \cite{Potera1997} \\
\end{longtable}

\newpage
\begin{longtable}{|p{4.6cm}|p{2.0cm}|p{2.4cm}|p{2.8cm}|p{4.4cm}|}
\caption{\label{tab:Features} Features used for predicting the target variables. AA - Angler's Atlas database, StatCan - Statistics Canada, AB - Alberta, BC - British Columbia, ON - Ontario}\\
\hline
\textbf{Feature} & \textbf{Data type} & \textbf{Temporal aggregation} & \textbf{Dimensionality} & \textbf{Source} \\ \hline
\endfirsthead
\multicolumn{5}{c}{{\tablename\ \thetable{} -- continued from previous page}} \\
\hline
\textbf{Feature} & \textbf{Data Type} & \textbf{Temporal aggregation} & \textbf{Dimensionality} & \textbf{Source} \\ \hline
\endhead
\hline \multicolumn{5}{r}{{Continued on next page}} \\ \hline
\endfoot
\hline
\endlastfoot
Environment &  &  &  &  \\
\hline
Minimum air temperature [°C] & Numerical & Average & Spatiotemporal & BioSIM \\
Mean air temperature [°C] & Numerical & Average & Spatiotemporal & BioSIM \\
Maximum air temperature [°C] & Numerical & Average & Spatiotemporal & BioSIM \\
Total precipitation [mm] & Numerical & Average & Spatiotemporal & BioSIM \\
Dew point temperature [°C] & Numerical & Average & Spatiotemporal & BioSIM \\
Relative humidity [\%] & Numerical & Average & Spatiotemporal & BioSIM \\
Solar radiation [watt/m2] & Numerical & Average & Spatiotemporal & BioSIM \\
Atmospheric pressure [hPa] & Numerical & Average & Spatiotemporal & BioSIM \\
Wind speed at 2 m [km/h] & Numerical & Average & Spatiotemporal & BioSIM \\
Degree days [°C] & Numerical & Average & Spatiotemporal & \\
River or lake & Boolean & - & Spatial & AA \\
Elevation [m] & Numerical & - & Spatial & BioSIM \\
Surface area [m2] & Numerical & - & Spatial & AA \\
Shoreline [m] & Numerical & - & Spatial & AA \\
\hline
Socioeconomics &  &  &  &  \\
\hline
Human population size in surrounding area [people] & Numerical & - & Spatial & StatCan (year 2021) \\
Mean income in surrounding area [CA\$] & Numerical & - & Spatial & StatCan (year 2021) \\
Median income in surrounding area [CA\$] & Numerical & - & Spatial & StatCan (year 2021) \\
Distance to next urban area [m] & Numerical & - & Spatial & AA, StatCan (year 2021) \\
Distance to road [m] & Numerical & - & Spatial & AA, StatCan (year 2021)\\
Campground availability & Boolean & - & Spatial, only available for BC & British Columbia Lodging \& Campgrounds Association \\
Change in work hours due to Covid-19 [\%] & Numerical & Avergae & Temporal (quarterly, from Q4 2019 to Q4 2021) & StatCan \\
Average hourly wages & Numerical & Avergae & Temporal (monthly), until January 2022 & StatCan \\
Consumer price index & Numerical & Avergae & Temporal (monthly, until January 2022) & StatCan \\
Covid cases in the last seven days & Numerical & Avergae & Spatiotemporal (Province) & \cite{Berry2021} \\
\hline
Fisheries management and events &  &  &  &  \\
\hline
Bag limitations & Boolean & - & Spatial & \cite{Ontario2019, Alberta2020, BC2019}\\
Fish size limitations & Boolean & - & Spatial & \cite{Ontario2019, Alberta2020, BC2019}\\
Catch-and-release regulation & Boolean & - & Spatial & \cite{Ontario2019, Alberta2020, BC2019}\\
Water body closure & Categorical & Sums in three variables & Spatiotemporal & \cite{Ontario2019, Alberta2020, BC2019}\\
Weekend day or weekday & Boolean & Removed & Temporal & - \\
Public holiday (+ connected weekend) & Boolean & Sum & Spatiotemporal & \url{https://www.statutoryholidays.com/}\\
Number of active fishing licenses in the province & Numerical & Sum & Spatiotemporal (Province-wide available for AB and BC) &  Government of Alberta; Freshwater Fisheries Society of BC\\
Stocking event in the year & Boolean & Sum & Spatiotemporal &  Ministry ON (April 1, 2021);  Government of Alberta: Fish Stocking Report 2022; Freshwater Fisheries Society of BC, Regional report (2022) \\ 
Weeks since the last stocking event [weeks] & Numerical & Average & Spatiotemporal & \\
Whirling disease detected & Boolean & - & Spatial (from August 2020), only available for AB & Alberta Environment and Parks, Open Government Licence – Alberta \\
\end{longtable}

\newpage
\begin{table}
    \caption{Accuracy scores of test sets of daily prediction models with additional features. Accuracy scores refer to predictions over different spatial extents: the mean (i) over all provinces (including predicting only rivers or only lakes), (ii) over single provinces, (iii) over single regions, and (iv) over single water bodies.}
    \label{tab:AdditionalFeatures}
    \centering
    \begin{tabular}{|l|l|r|r|} \hline 
         \textbf{Additional feature}&  \textbf{Spatial extent}& \textbf{Mean catch rate}& \textbf{Total fishing duration}\\ \hline 
    No additional feature & All provinces& 0.579 & 0.605\\
         & Province & 0.562 & 0.597\\  
         &  Region & 0.579 & 0.605\\ 
         &  Water body & 0.735 & 0.735\\ \hline 
    Julian day &  All provinces& 0.579 & 0.605\\
         & Province & 0.562 & 0.596\\ 
         &  Region & 0.579 & 0.605\\  
         &  Water body & 0.733 &0.732\\ \hline 
    Week number &  All provinces& 0.549 & 0.565\\
         & Province & 0.553 & 0.617\\ 
         &  Region & 0.549 & 0.565\\
         &  Water body & 0.660 & 0.659\\ \hline 
    Month& All provinces & 0.578 & 0.604\\
         & Province & 0.563 & 0.597\\
         & Region & 0.579 & 0.604\\ 
         & Water body & 0.721 & 0.724\\ \hline 
    Website views in  & All provinces & 0.582 & 0.626 \\
    previous week     & Province & 0.564 & 0.606\\
         & Region & 0.582 & 0.626\\ 
         & Water body & 0.710 & 0.723\\ \hline
    \end{tabular}
\end{table}

\clearpage
\printbibliography

\end{document}